\documentclass[preprint]{elsarticle}
\journal{European Journal of Control}

\usepackage{amsmath,amssymb,amsfonts}
\usepackage{algorithmic}
\usepackage{graphicx}
\usepackage{textcomp}
\usepackage{breqn}
\usepackage{comment}
\usepackage{mathtools}
\usepackage{graphics}

\usepackage{xcolor}

\DeclareMathOperator*{\argmin}{arg\,min}

\newtheorem{thm}{Theorem}[section]

\newtheorem{assum}[thm]{Assumption}
\newtheorem{prop}[thm]{Proposition}

\newproof{pf}{Proof}
\newtheorem{defn}[thm]{Definition}

\newdefinition{rem}[thm]{Remark}

\DeclareMathOperator*{\minimize}{minimize}

\begin{document}
\begin{frontmatter}
\title{Multi-objective low-thrust spacecraft trajectory design using reachability analysis}
\author[Oxford]{Nikolaus Vertovec}
\ead{nikolaus.vertovec@eng.ox.ac.uk}
\author[Paderborn]{Sina Ober-Bl\"obaum}
\ead{sinaober@math.uni-paderborn.de}
\author[Oxford]{Kostas Margellos}
\ead{kostas.margellos@eng.ox.ac.uk}
\address[Oxford]{Department of Engineering Science, University of Oxford, Oxford, OX1 3PJ, United Kingdom}
\address[Paderborn]{Department of Mathematics, Paderborn University,
33100, Paderborn, Germany}
\begin{abstract}
    One of the fundamental problems in spacecraft trajectory design is finding the optimal transfer trajectory that minimizes the propellant consumption and transfer time simultaneously. We formulate this as a multi-objective optimal control (MOC) problem that involves optimizing over the initial or final state, subject to state constraints. Drawing on recent developments in reachability analysis subject to state constraints, we show that the proposed MOC problem can be stated as an optimization problem subject to a constraint that involves the sub-level set of the viscosity solution of a quasi-variational inequality. We then generalize this approach to account for more general optimal control problems in Bolza form. We relate these problems to the Pareto front of the developed multi-objective programs. The proposed approach is demonstrated on two low-thrust orbital transfer problems around a rotating asteroid.
\end{abstract}
\end{frontmatter}
\section{Introduction}
\label{sec:introduction}
Since the Galileo mission in 1991 we have seen a steady increase in proposed missions to asteroids and comets, as they might hold the key to many scientific questions including the origins of life on earth \cite{Antreasian1994}. The Dawn mission to Vesta and Ceres proved the viability of low-thrust electric propulsion for asteroid exploration \cite{Abrahamson2013, Parcher2011}, and it is expected that many upcoming missions will rely on similar low-thrust propulsion. While there has been a significant study of interplanetary transfer trajectories using low-thrust propulsion, comparatively little research has been conducted on the trajectory design in the vicinity of asteroids. We investigate a spacecraft trajectory design problem around an asteroid, where the objective is to use minimal amounts of propellant to raise an orbit while keeping flight times as short as possible. This is a multi-objective optimal control (MOC) problem, whereby one seeks to find the optimal way a dynamical system can perform a certain task, while minimizing or maximizing a set of, usually contradictory and incommensurable, objective functions \cite{OberBlobaum2012}. \textcolor{black}{Conventional optimization techniques for spacecraft trajectory design often fall into two categories, indirect methods, based on the calculus of variations, and direct methods, whereby the optimal control problem is reformulated as a nonlinear program. Direct approaches rely on parametrization and while a candidate solution is found, there are no guarantees on the optimality of the solution. Indirect methods, meanwhile rely on necessary analytic conditions for optimality using Lagrange multipliers. Yet while optimality of the obtained solutions may be guaranteed, indirect approaches, such as multiple shooting methods, rely on a good initial approximation of the optimal trajectory \cite{SHIRAZI2018, VonStryk1992}. A third approach is dynamic programming, whereby the optimality conditions are formulated in continuous time based on the Hamilton-Jacobi-Bellman (HJB) equation \cite{SHIRAZI2018}, however, it is hampered by the so-called curse of dimensionality. Despite this, unlike direct approaches, optimality is guaranteed, and unlike indirect approaches, the solution does not rely on an initial approximation of the optimal trajectory. We expand on this third approach by taking advantage of recent developments in reachability analysis.} 

Reachability analysis aims to find the set of points from which a target can be reached within a given time, subject to constraints. It forms a fundamental part of the dynamics and control literature and has been used extensively for controller synthesis of complex systems \cite{Aubin2002, Lygeros1999, Mitchell2005a}. In recent years we have seen considerable research being conducted into computing reachable sets using Hamilton-Jacobi (HJ) reachability analysis, whereby the reachable set is derived from the viscosity solution of a HJB equation accounting also for the presence of state constraints. \textcolor{black}{Such a HJB framework is presented in \cite{Bokanowski2010, Margellos2011} with more general value problems bypassing previous regularity issues presented in \cite{Altarovici2013}. In \cite{Fisac2015} an extension of the HJB framework to time-varying targets and constraints is considered and in \cite{Desilles2019} the approach is extended to multi-objective control problems. HJ reachability has also been successfully applied to various aerospace applications including air traffic control \cite{Margellos2013}, the climbing problem of multi-stage launchers \cite{Bokanowski2015}, payload optimization \cite{Bokanowski2017}, as well as most recently to the complete model of the ascent problem of multi-stage launchers \cite{Bokanowski2018}.} One of the advantages of using HJ reachability is that the optimal trajectory can easily be constructed once the reachable set has been computed. This makes HJ reachability attractive for problems that require computing trajectories for various different initial states. 

For the spacecraft trajectory design problem considered in this paper, there are two possible formulations for minimizing the burnt propellant. The first assumes that the initial mass is a free optimization variable. This approach is common during mission design where the total required fuel budget is being calculated. To this end, we formulate the spacecraft trajectory design problem as a MOC problem and show that it can be equivalently stated as an optimization problem subject to a constraint that involves the sub-level set of a certain value function. The latter is shown to be the unique continuous viscosity solution of a quasi-variational inequality that involves a HJB equation. Such value functions have been defined in \cite{Bokanowski2010, Margellos2011} to account for the presence of state constraints. This formulation allows characterizing the Pareto front of the formulated MOC problem and also facilitates its computation by means of available numerical tools.

The second formulation of the spacecraft trajectory design problem assumes a fixed initial mass and the objective is to maximize the remaining mass after completing a given orbital maneuver. This approach is more common when the maneuver needs to be added to a given mission and the available fuel is non-negotiable. This formulation had been previously investigated by the authors in \cite{Vertovec2021}. To solve this second formulation of the spacecraft trajectory design problem, we draw on research from \cite{Desilles2019} and extend our formulation of the MOC problem to introduce an auxiliary state, allowing us to account for arbitrary problems in Bolza form, and, together with appropriate normalization and approximations allowing for a reduction of the state space, greatly improving on the method presented in \cite{Vertovec2021}.

Thus our contributions can be summarized as
\begin{enumerate}
    \item the formulation of an efficient constrained MOC problem for low-thrust spacecraft trajectory design that optimizes only over the set of admissible initial states and transition times,
    \item the reduction of the state space through the use of appropriate approximations,
    \item the expansion of the MOC problem to allow for a generalization of the proposed methodology for arbitrary multi-objective problems in Bolza form.
\end{enumerate}

This paper is organized into six sections. Section \ref{sec:Mathematical_description_and_physical_modeling} contains details regarding the derivations of the spacecraft dynamics as well as the definitions of the constraints pertaining to its behavior. In Section \ref{sec:problem_statement} the optimal control problem is formulated while Section \ref{sec:Solution_to_multi-objective_optimal_control_problems} describes how the set of admissible initial states is derived from the viscosity solution of a quasi-variational inequality. Section \ref{sec:numerics} is dedicated to the numerical computation and case study of an orbital transfer around a rotating asteroid. Finally, Section \ref{sec:conclusion} provides concluding remarks and directions for future work.

\section{Mathematical description and physical modeling} \label{sec:Mathematical_description_and_physical_modeling}
\subsection{Spacecraft equations of motion}
We begin by modeling the dynamics of the spacecraft. \color{black}The spacecraft thrust is defined in spherical coordinates as 
\begin{equation}
	\mathbf{u}(t) \coloneqq [\boldsymbol\alpha(t), \boldsymbol\delta(t), \textbf{T}(t)] \in \mathcal{U},
\end{equation}
where $\boldsymbol\alpha(t) \in  [-\pi, \pi]$ is the incidence angle, $\boldsymbol\delta(t) \in [-\frac{\pi}{2}, \frac{\pi}{2}]$ is the sideslip angle and $\textbf{T}(t) \in [0, T_{\max}]$ is the variable thrust, with $T_{\max}$ denoting the maximal allowable thrust. The Cartesian transformation of the thrust vector is denoted by $\mathbf{u}_x(t)$,  $\mathbf{u}_y(t)$ and $\mathbf{u}_z(t)$, respectively. The compact set $\mathcal{U} = [-\pi, \pi] \times [-\frac{\pi}{2}, \frac{\pi}{2}] \times [0, T_{\max}]$ is the set of possible control input values while $\mathbf{u} \in \mathcal{U}_{ad}$ denotes the control policy and $\mathcal{U}_{ad}$ denotes the set of admissible policies which is the set of Lebesgue time measurable functions from $[-\infty, 0]$ to $\mathcal{U}$. Note that time here is considered to be non-positive to facilitate the reachability problem exposition in Section \ref{sec:problem_statement}. \color{black}Boldface notation is used to denote time varying functions such as trajectories and policies, while non-boldface notation is used to denote scalars and vectors.
The equations of motion of the spacecraft around a rotating body can be expressed in 3-dimensional Euclidean space as a second-order ordinary differential equation (see eg., \cite{Jiang2014})
\begin{multline}
	2 \mathbf{\Omega}(t) \times \frac{d \textbf{R}(t)}{d t}+\mathbf{\Omega}(t) \times (\mathbf{\Omega}(t) \times \textbf{R}(t))+\frac{d \mathrm{U}(\textbf{R}(t))}{d \textbf{R}} \\
	+ \frac{d \mathbf{\Omega}(t)}{d t} \times \textbf{R}(t) -\frac{\textbf{u}(t)}{m(t)} = -\frac{d^2 \textbf{R}(t)}{d t^2},
	\label{eqn: system_dynamics}
\end{multline}
where \textbf{R}(t) is the radius vector from the asteroid's center of mass to the particle, the first and second time derivatives of \textbf{R}(t) are with respect to the body-fixed coordinate system, $\mathrm{U}(\textbf{R}(t))$ is the gravitational potential of the asteroid and $\Omega$ is the rotational angular velocity vector of the asteroid relative to inertial space. The term $2 \mathbf{\Omega}(t) \times \frac{d \textbf{R}(t)}{d t}$ describes the Coriolis forces, $\mathbf{\Omega}(t) \times (\mathbf{\Omega}(t) \times \textbf{R}(t))$, the centrifugal forces and $\frac{d \mathbf{\Omega}(t)}{d t} \times \textbf{R}(t)$ the Euler forces. We consider an asteroid rotating uniformly with constant magnitude $\omega$ around the z-axis. Therefore, the Euler forces can be neglected and we can express the rotation vector as $\Omega \coloneqq \omega e_z$, where $e_z$ is the unit vector along the z-axis. Following \cite{Greenwood1988}, the radius vector and its derivatives are given by 
\begin{equation}
	\mathbf{R}(t) \coloneqq \begin{bmatrix} \mathbf{x}(t) \\ \mathbf{y}(t) \\ \mathbf{z}(t) \end{bmatrix}, \quad 
	\frac{d \textbf{R}(t)}{d t} = \begin{bmatrix} \mathbf{v}_x(t) \\ \mathbf{v}_y(t) \\\mathbf{v}_z(t) \end{bmatrix}.
\end{equation}
The Coriolis and centrifugal forces (the first two terms in \eqref{eqn: system_dynamics}) acting on the spacecraft are thus
\begin{align}
	2  \Omega \times \frac{d \textbf{R}(t)}{d t} = \begin{bmatrix} -2 \: \mathbf{v}_y(t) \\
			2 \omega \textbf{v}_x(t) \\
			0\end{bmatrix}, \\
	\Omega \times (\Omega \times \textbf{R}(t)) =  \begin{bmatrix} -\omega ^2 \mathbf{x}(t) \\
			-\omega ^2 \mathbf{y}(t)                             \\
			0 \end{bmatrix}.
\end{align}
To model the current position, velocity, and available propellant, we define the state vector 
\begin{equation}
	r \coloneqq \begin{bmatrix} x, y, z, v_x, v_y, v_z, \Delta m \end{bmatrix}^T \in \mathbb{R}^7,
\end{equation}
\textcolor{black}{where $\Delta m \in \mathbb{R}_{\geq 0}$ denotes the available propellant. The total spacecraft mass can be expressed as $m(t) = m_0 + \Delta m(t)$, where $m_0$ denotes the dry mass of the spacecraft.}
Following our derivations from \eqref{eqn: system_dynamics}, we can formulate the dynamics of the spacecraft, $\dot{r} = f(r, u)$, as
\begin{equation}
	f(r, u) = \begin{bmatrix}
		v_x \\ v_y \\ v_z  \\
		U_x(x,y,z) + \omega ^2 x  + 2 \omega v_y  + \frac{u_x}{m_0+\Delta m}                \\
		U_y(x,y,z) + \omega ^2 y  - 2 \omega v_x + \frac{u_y}{m_0+\Delta m}      \\
		U_z(x,y,z) + \frac{u_z}{m_0+\Delta m} \\
		-\frac{\sqrt{u_x^2 + u_y^2 + u_z^2}}{v_{\mathrm{exhaust}}}\end{bmatrix},
	\label{eqn:dynamics}
\end{equation}
where $v_{\mathrm{exhaust}} \in \mathbb{R}_{\geq 0}$ is the exhaust velocity used to express the depletion of mass as propellant is burned, $U_x$, $U_y$ and $U_z$ are the derivatives of the gravitational potential in the direction of the unit vectors $e_x$, $e_y$ and $e_z$, respectively, and where for brevity we neglect the time dependence by denoting $r=\textbf{r}(t)$, $v_x = \textbf{v}_x(t)$, and similarly for the other states.

\subsection{State constraints}
Since the dynamics of the spacecraft were derived for orbits in the vicinity of the asteroid, we need to enforce state constraints on $x, y, z$. We naturally also need to ensure that we bound the amount of propellant available. Assuming that the burnout mass of the spacecraft is the same as the dry mass, we set $m_{\min} \coloneqq 0$ and $m_{\max} \coloneqq m_{\mathrm{propellant}}$ and impose $m_{\min} \leq \Delta m \leq m_{\max}$.

Due to particles ejected from the asteroid, we do not want to fall below a circular orbit with radius $\rho \coloneqq \sqrt{x^2 + y^2 + z^2}$ of approximately $\rho_{\min} = 1$ km. \textcolor{black}{Furthermore, in order for the two-body problem under discussion to be valid and the influence of other bodies in the solar system to be negligible, we need to stay within the sphere of influence (SOI) of the asteroid. The SOI can be approximated as in \cite{Seefelder2002} by $\rho_{SOI} \approx a \left( \frac{M_1}{M_2}\right)^{\frac{2}{5}}$}, where $a$ is the semi-major axis of the asteroid's orbit around the sun ($1.5907 \cdot 10^{8}$ km), $M_1$ is the Mass of the asteroid ($1.4091 \cdot 10^{12}$ kg) and $M_2$ is the mass of the sun ($1.9890 \cdot 10^{30}$ kg). Therefore, the sphere of influence of the asteroid is approximately $\rho_{\max} = \rho_{SOI} \approx 8.74$ km. The set of states that satisfy the aforementioned restrictions is given by
\begin{equation*}
\mathcal{K} \coloneqq \left\{ r \in \mathbb{R}^7 :  \rho \in [\rho_{\min}, \rho_{\max}], m \in [m_{\min}, m_{\max}]\right\}.
\end{equation*} 
The target orbit that we would like to transfer to is denoted by the closed target set $\mathcal{C} \subset \mathcal{K}$.

The initial orbit that we start at is denoted by the closed initial set $\mathcal{I} \subset \mathcal{K}$. Note that the initial and the target orbit restrict only the position and the velocity, but allow the mass to take any admissible value within $[m_{\min}, m_{\max}]$.

While Cartesian coordinates are useful for modeling the behavior of an object around a rotating body, since we restrict all admissible states to lie within the set $\mathcal{K}$, which constrains the radius $\rho$, it is more efficient to recast our problem in spherical coordinates. To this end, define $ a_\rho, a_\theta, a_\psi$ as the transformations of
\begin{align*}
a_x \coloneqq &{} U_x(x,y,z) + \omega ^2 x  + 2 \omega v_y, \\
a_y \coloneqq & U_y(x,y,z) + \omega ^2 y  - 2 \omega v_x, \\
a_z \coloneqq& U_z(x,y,z).
\end{align*}
The tangential velocity in the x-y plane, $v_t$, and its perpendicular counterpart, $v_\perp$, can then be defined as follows:
\begin{align}
    \begin{bmatrix}
    	v_t \\
    	v_\perp
    \end{bmatrix} = &{} \begin{bmatrix}
    	\rho \dot{\theta}  \sin{\psi}  \\
    	\rho \dot{\psi} 
    \end{bmatrix}, \\
    \begin{bmatrix}
    	a_t \\
    	a_\perp
    \end{bmatrix} = & \begin{bmatrix}
    	\sin{\psi} [a_\theta \rho +  \dot{\theta} v_\rho] + \dot{\theta} v_\perp \cos{\psi}\\
    	\rho \dot{\psi}+ a_\psi \rho
    \end{bmatrix}.
\end{align}
Then we can restate the system dynamics in spherical coordinates as
\begin{align}
	r  & = \begin{bmatrix}
		\rho,
		\theta,
		\psi,
		v_\rho,
		v_t,
		v_\perp,
		\Delta m\end{bmatrix}^T \in \mathbb{R}^7, \\
	f(r,u)  & = \begin{bmatrix}
		v_\rho\\ 
		\frac{v_t}{\rho \sin{\psi}} \\ 
		\frac{v_\perp}{\rho}  \\
		a_\rho  + \frac{T}{m_0+\Delta m} \cos{\alpha}  \\
		a_t  + \frac{T}{m_0+\Delta m} \sin{\alpha} \sin{\delta} \\
		a_\perp + \frac{T}{m_0+\Delta m} \sin{\alpha} \cos{\delta} \\
		- \frac{T}{v_{\mathrm{exhaust}}}\end{bmatrix}.
		\label{eqn:sph-dynamics}
\end{align}
 With a slight abuse of notation, we also redefine $\mathcal{I}$, $\mathcal{C}$, and $\mathcal{K}$ in spherical coordinates. \color{black}Finally, we impose the following assumptions on the spacecraft dynamics.
\begin{assum}
	For every $r \in \mathcal{K}$ the set $\Big\{f(r,u) : u \in \mathcal{U} \Big\}$ is a compact convex subset of $\mathbb{R}^{7}$.
	\label{assu:compact_convex}
\end{assum}
\begin{assum}
	$f:\mathbb{R}^{7} \times \mathcal{U} \rightarrow \mathbb{R}^{7}$ is bounded and there exists an $L_{f} > 0$ such that for every $ u_1, u_2 \in \mathcal{U}$,
	\begin{equation*}
		|| f(r_{1}, u_1) - f(r_{2}, u_2) || \leq L_{f} ||r_{1} - r_{2}||.
	\end{equation*}
	\label{assu:Lipschitz}
\end{assum}
Due to norm equivalence, the choice of norm is irrelevant and not further discussed. Using Assumptions \ref{assu:compact_convex} and \ref{assu:Lipschitz}, for any control policy $\textbf{u} \in \mathcal{U}_{ad}$, any initial state $r_0 \in \mathcal{K}$ and transfer time $t_f > 0$, the system admits a unique, absolutely continuous solution on $[-t_f, 0]$ (see \cite{Sastry1999}).
\color{black}

\section{Problem Statement} \label{sec:problem_statement}
\subsection{Multi-objective optimal control problem}
Having defined the system dynamics, we are now in a position to discuss how to find trajectories that start on an initial orbit, $\mathcal{I}$, and take the spacecraft to some final orbit, $\mathcal{C}$. Additionally, the objective is to keep the flight time and required propellant as small as possible. Thus, the multi-objective optimal control problem can be formulated as a minimization problem whereby the first goal is to minimize the required propellant, $\Delta m$, and the second is to minimize the required time for the orbit change, i.e., the transfer time, denoted by $t_f$.
The trajectory, $\mathbf{r}$, which is the solution of \eqref{eqn:sph-dynamics}, belongs to the Sobolev space $\mathbb{W}^{1,1}(\mathbb{R}^7)$. The set of trajectory-control pairs on $[-t_f, 0]$ starting at $r_0$ with transfer time $t_f$ is denoted as:
\color{black}
\begin{multline*}
	\Pi_{r_0,t_f} \coloneqq \big\{(\mathbf{r},\mathbf{u}) : \dot{\mathbf{r}}(t) = f(\mathbf{r}(t),\mathbf{u}(t)), \quad \forall t \in [-t_f, 0]; \\
	\mathbf{r}(-t_f) = r_0 \big\} \subset \mathbb{W}^{1,1}(\mathbb{R}^7) \times \mathcal{U}_{ad}.
\end{multline*}
\color{black}
Note that as in \cite{Chen2018} we adopt the convention that $0$ denotes the terminal time hence the transfer time $t_f$ denotes the time duration. \color{black}Under Assumption \ref{assu:compact_convex} and by Filippov's Theorem \cite[pg.121]{Liberzon2011}, we can conclude, that $\Pi_{r_0,t_f}$ is compact. 
\begin{rem}
    For similar applications as the one discussed in this paper, where Assumption \ref{assu:compact_convex} might not hold, we refer to \cite{Bokanowski2018} where convexification of the dynamics is considered in order to ensure that the set of absolutely continuous solutions of the problem is closed.
\end{rem}
\color{black}
The set of admissible (in the sense of satisfying the state constraints) trajectory-control pairs on $[-t_f, 0]$ starting at $r_0$ with transfer time $t_f$ is denoted as:
\color{black}
\begin{multline*}
	\Pi_{r_0,t_f}^{\mathcal{K}, \mathcal{C}} \coloneqq \big\{(\mathbf{r},\mathbf{u}) \in \Pi_{r_0,t_f} : \mathbf{r}(t) \in \mathcal{K}, \quad \forall t \in [-t_f, 0]; \\
	\mathbf{r}(0) \in \mathcal{C} \big\} \subset \mathbb{W}^{1,1}(\mathbb{R}^7) \times \mathcal{U}_{ad}.
\end{multline*}
\color{black}
Finally, the set of admissible initial state and transfer time pairs is denoted as 
\begin{equation*}
	\Gamma\coloneqq \big\{(r_0, t_f) \in \mathbb{R}^7 \times [0,+\infty) \quad \text{such that} \quad \Pi_{r_0,t_f}^{\mathcal{K}, \mathcal{C}} \neq \emptyset \big\}.
\end{equation*}
For a given initial state $r_0 \in \mathbb{R}^7$ and transfer time $t_f \in [0, +\infty)$, we can define the cost functions as $J_1(r_0, t_f) \coloneqq \Delta m$ and $J_2(r_0, t_f) \coloneqq t_f$, where $\Delta m$ is the $7$-th element of the state vector $r_0$. The 2-dimensional objective function $J : \mathbb{R}^7 \times [0,+\infty) \rightarrow \mathbb{R}^2$ can then be written as
\begin{equation}
	J(r_0, t_f) \coloneqq \left[ J_1(r_0, t_f), J_2(r_0, t_f) \right]^T.
\end{equation}
We are now in a position to formulate the multi-objective optimal control problem under study as 
\begin{equation}
\begin{aligned}
& \minimize_{(r_0, t_f) \in \mathcal{I} \times [0, \infty)}
& J(r_0, t_f) \\
& \text{subject to} & (r_0, t_f) \in \Gamma
\end{aligned}
	\label{eqn:MOC Problem}
\end{equation}
\subsection{Pareto optimality}
The solution of \eqref{eqn:MOC Problem} in general does not consist of a single isolated point, but rather a set of optimal compromises between the objectives $J_1$ and $J_2$ \cite{Miettinen1998}.
\begin{defn}
A solution $(r_0, t_f)$ is considered Pareto optimal if $\nexists (\hat{r_0}, \hat{t}_f) \in \Gamma$ such that $J(\hat{r_0}, \hat{t}_f) < J(r_0, t_f)$,	
\label{def:Pareto Optimality}
\end{defn}
where a vector $a$ is considered less than $b$ (denoted $a < b$) if for every element $a_i$ and $b_i$ the relation $a_i < b_i$ holds. The relations $\leq, \geq, >$ are defined in an analogous way. 
Following Definition \ref{def:Pareto Optimality}, a solution $(r_0, t_f)$ is considered Pareto optimal if it is not possible to improve all its performance metrics $J_1(r_0, t_f), J_2(r_0, t_f)$ simultaneously. The set of Pareto optimal solutions is called the Pareto set $\mathcal{P}_S$, while its image is the Pareto front $\mathcal{P}_F$. \textcolor{black}{Therefore, the solution of \eqref{eqn:MOC Problem}, i.e., the set of minimizing $(r_0, t_f)$ pairs, is the desired Pareto set, while the cost function corresponding to the minimizing $(r_0, t_f)$ pairs is the Pareto front.}

To allow for mission designers to determine a compromise between minimizing required propellant and transfer times, we wish to compute the Pareto front. However, the unconventional constraint in \eqref{eqn:MOC Problem} ensuring a solution $(r_0, t_f)$ is feasible, prevents us from solving \eqref{eqn:MOC Problem} with standard MOC problem solvers. Therefore, we will next discuss how we can recast the constraint $(r_0, t_f) \in \Gamma$ to a standard nonlinear inequality constraint, which will allow us to compute the Pareto front by means of conventional MOC problem solvers.

\section{Solution to multi-objective optimal control problems} \label{sec:Solution_to_multi-objective_optimal_control_problems}
To find an equivalent formulation for the constraint $(r_0, t_f) \in \Gamma$, let $g(r)$ and $\nu (r)$ be two Lipschitz functions (with Lipschitz constants $L_g$ and $L_{\nu}$, respectively) chosen such that
\begin{align*}
	g(r) \leq &{} 0 \iff r \in \mathcal{K}, \\
	\nu(r) \leq & 0 \iff r \in \mathcal{C}.
\end{align*}
This can be achieved by choosing $g(r)$ and $\nu(r)$ as the signed distance to the set $\mathcal{K}$ and $\mathcal{C}$, respectively. 

Next, we consider the value function $\omega$:
\begin{equation}
	\omega(r_0, t_f) \coloneqq \inf_{(\mathbf{r}, \textbf{u}) \in \Pi_{r_0, t_f}} \Big\{ 
	\nu(\mathbf{r}(0))
	\bigvee
	\max_{\tau \in [-t_f, 0]} g(\mathbf{r}(\tau))
	\Big\},
	\label{eqn:valuefuncBRS}
\end{equation} 
where $a \bigvee b$ denotes $\max(a,b)$. 
We are now in a position to use the value function to decide if, for a given initial state and transfer time, there exists a corresponding admissible trajectory. Thus we can introduce an equivalent formulation of \eqref{eqn:MOC Problem}.
\begin{thm}
The constrained MOC problem, \eqref{eqn:MOC Problem}, is equivalent to
\label{thm:newMoc}
\end{thm}
\begin{equation}
    \begin{aligned}
    & \minimize_{(r_0, t_f) \in \mathcal{I} \times [0, \infty)}
    & J(r_0, t_f) \\
    & \text{subject to}
    & \omega(r_0, t_f) \leq 0.
    \end{aligned}
    \label{eqn:newMoc}
\end{equation}
\begin{pf}
    We show that  $(r_0, t_f) \in \Gamma \iff \omega(r_0, t_f) \leq 0$.
    
    \textit{Case A}:
    Consider $(r_0, t_f) \in \Gamma$. For the sake of contradiction assume that $\omega(r_0, t_f) > 0$. This then implies that for all $(\mathbf{r},\mathbf{u}) \in \Pi_{r_0,t_f}^{\mathcal{K}, \mathcal{C}}$ either $\nu(\mathbf{r}(0)) > 0 \iff \mathbf{r}(0) \notin \mathcal{C}$ or there exists $\tau \in [-t_f, 0]$ such that $g(\mathbf{r}(t)) > 0 \iff \mathbf{r}(t) \notin \mathcal{K}$. This contradicts the fact that $(r_0, t_f) \in \Gamma$ establishing that $(r_0, t_f) \in \Gamma$ implies $\omega(r_0, t_f) \leq 0$
    
    \textit{Case B}:
    Consider $(r_0, t_f) \in \mathcal{I} \times [0, \infty)$, such that $\omega(r_0, t_f) \leq 0$. Under Assumption \ref{assu:compact_convex}, applying Weierstrass' Theorem on the existence of minima for compact sets \cite{Altarovici2013, Bardi1997}, we can conclude that the infimum over $\Pi_{r_0,t_f}$ exists, and thus, $\omega(r_0, t_f) \leq 0$ implies the existence of a trajectory-control pair $(\mathbf{r},\mathbf{u}) \in \Pi_{r_0,t_f}$, such that for all $t \in [-t_f, 0], \: g(\mathbf{r}(t)) \leq 0$ and $\nu(\mathbf{r}(0)) \leq 0$. By definition of the function $g$ and $\nu$, we thus have $\mathbf{r}(t) \in \mathcal{K}$ for all $t \in [-t_f, 0]$ and $\mathbf{r}(0) \in \mathcal{C}$, which in turns implies $(\mathbf{r},\mathbf{u}) \in \Pi_{r_0,t_f}^{\mathcal{K}, \mathcal{C}}$. Therefore, 
    \begin{equation*}
        \omega(r_0, t_f) \leq 0 \Rightarrow (r_0, t_f) \in \Gamma,
    \end{equation*}
   thus concluding the proof.
\end{pf}
Theorem \ref{thm:newMoc} implies that the Pareto front can be computed from the solution of \eqref{eqn:newMoc}. To achieve this we discuss how to compute $\omega$. 

\subsection{Value function computation}
To begin to discuss how $\omega$ can be obtained, we introduce the Hamiltonian $ H: \mathbb{R}^7 \times \mathbb{R}^7 \rightarrow \mathbb{R}$,
\begin{equation}
	H(r, q) \coloneqq  -\min_{u \in \mathcal{U}} \left(q^T f(r,u)\right),
\end{equation}
where $q \in \mathbb{R}^7$ is the costate vector.
\begin{thm}
	 The value function $\omega$ is the unique continuous viscosity solution of the following quasi-variational inequality
	\label{thm:HJB}
\end{thm}
\begin{equation*}
    	\begin{cases}
    		\begin{aligned}
    			0 = \max \big\{g(r)-\omega(r, t), \partial_{t} \omega + H(r, \nabla_{r} \omega)\big\}\\
    			\text{for all} \quad t \in [0, \infty), \: r \in \mathbb{R}^7,
    		\end{aligned} \\		
    		\begin{aligned}
    			\omega(r, 0) = {} &
    			\left(\nu(r)
    			\bigvee
    				g(r)
    			\right) \quad \text{for all} \quad r \in \mathbb{R}^7,
    		\end{aligned}
    	\end{cases}
	\label{eqn:HJB}
\end{equation*}
Since the Dynamic Programming Principle \cite{Bardi1997} holds for $h \in [0, t_f] $, $(r_0, t_f) \in \mathcal{K} \times \mathbb{R}$ with $t_f \geq 0$: 
\begin{equation*}
	\omega(r_0, t_f) = \inf_{(\mathbf{r}, \textbf{u}) \in \Pi_{r_0, t_f}} \Big\{
	\omega(\mathbf{r}(h-t_f), t_f-h)
	\bigvee \max_{s \in [-t_f, h-t_f]} g(\mathbf{r}(s))
	\Big\},
\end{equation*}
the proof of Theorem \ref{thm:HJB} follows standard arguments for viscosity solutions, as shown in \cite{Altarovici2013, Margellos2011}. Note that the infimum should be understood to be over the restriction of $\Pi_{r_0, t_f}$ over $[-t_f, h - t_f]$.

In order to solve the quasi-variational inequality in Theorem \ref{thm:HJB}, we employ a finite differences scheme. \textcolor{black}{As in \cite{Bokanowski2010, Margellos2011}, a consequence following from Theorem \ref{thm:HJB} is the Lipschitz continuity of the value function.}

\begin{prop}
	The value function $\omega$ is Lipschitz continuous.
	\label{prop:omega-lips}
\end{prop}
The proof is similar to the more general proof of Proposition \ref{prop:vartheta-lips} that is introduced in the sequel. \textcolor{black}{Proposition \ref{prop:omega-lips} allows us to make statements about the discrete-continuous error estimate, for which we refer to Theorem 5 in \cite{Bokanowski2010}, as well as the convergence of the value function, for which we refer to Proposition 6 in \cite{Fialho1999}.}

The Hamiltonian admits an explicit form. To this end, consider the term
\begin{equation*}
	C(r,q) \coloneqq
		q_1 v_\rho +
		q_2 v_t +
		q_3 v_\perp +
		q_4 a_\rho +
		q_5 a_t +
		q_6 a_\perp.
\end{equation*}
Then we can write the Hamiltonian as
\begin{multline}	
	H(r, q) \coloneqq - \min_{u \in \mathcal{U}} \Big(
	\frac{T}{m_0+\Delta m} \big(
	q_4 \cos{\alpha} \\
	+ \sin{\alpha} \left(
		q_5   \sin{\delta} +
		q_6   \cos{\delta}
		\right) 
	\big)
	- q_7 \frac{T}{v_{\mathrm{exhaust}}} 
	\Big) - C(r,q).
 \label{eqn:Hamiltonian_bfr_optangle}
\end{multline}
\textcolor{black}{As $T$ is always positive, the thrust angles can be optimized separately; see Appendix for more details. To this end, the derivation of $\alpha^*$ and $\delta^*$ follows a similar procedure as in \cite{Bokanowski2018} and is listed in the Appendix. After applying the optimal thrust angles, the Hamiltonian becomes affine in $T$ and we are able to find the optimal thrust magnitude. In particular, $H(r,q)$ becomes}
\begin{equation}
	H(r, q) = - \min_{T \in [0, T_{\max}]} \Big(
	-\frac{T}{m_0+\Delta m} \sqrt{q_4^2 + q_5^2 + q_6^2}
	- q_7 \frac{T}{v_{\mathrm{exhaust}}} 
	\Big) - C(r,q)
 \label{eqn:Hamiltonian_aftr_optangle}
\end{equation}
\begin{equation}
	\Rightarrow T^* \coloneqq \begin{cases}
		T_{\max}	 \quad \text{if} \: \frac{q_7}{v_{\mathrm{exhaust}}} + \frac{\sqrt{q_4^2 + q_5^2 + q_6^2}}{m_0 + \Delta m} \geq  0\\
		0 	\quad \text{otherwise}
	\end{cases}.
\end{equation}
\begin{rem}
    The singular control case, $\frac{q_7}{v_{\mathrm{exhaust}}} + \frac{\sqrt{q_4^2 + q_5^2 + q_6^2}}{m_0 + \Delta m} = 0$, is negligible for the consideration of the optimal Hamiltonian. For the optimal control computation during the trajectory calculation, we have numerically investigated the occurrence of singular arcs and found no instances in which $\frac{\sqrt{q_4^2 + q_5^2 + q_6^2}}{m_0 + \Delta m} = -\frac{q_7}{v_{\mathrm{exhaust}}}$ over an extended interval, thus we do not further consider the singular control case.
\end{rem}
Finally, applying $T^*$ and rewriting the minimum as the maximum of the negation of the associated function, the Hamiltonian takes the following analytic form
\begin{equation*}
	H(r, q) =  -C(r,q) + 
	 \max{\Big(
		q_7 \frac{T_{\max}}{v_{\mathrm{exhaust}}} +
		\frac{T_{\max}}{m_0+\Delta m} \sqrt{q_4^2 + q_5^2 + q_6^2}, 0
		\Big)}.
\end{equation*}
\subsection{Extension to problems in Bolza form}
We will now generalize our approach to problems where the objective functions do not rely only on the initial state and are written in Bolza form. To achieve this we need to introduce auxiliary states. \textcolor{black}{As in \cite{Altarovici2013, Desilles2019}, we show how problems in Bolza form are reformulated into Mayer form, and then show how problems in Mayer form are solved in a similar fashion as in Section \ref{sec:problem_statement}.}
Consider the $p$-dimensional objective function defined as:
\begin{equation}
	J_{\mathrm{Bolza}}(\mathbf{r}, \mathbf{u}, t_f) \coloneqq J_{\mathrm{t}}(\mathbf{r}(0))+ \\ \int_{-t_f}^{0}  J_{\mathrm{r}}(\mathbf{r}(s), \mathbf{u}(s)) ds,
\end{equation}
where $J_{\mathrm{t}}$ denotes the terminal cost and $J_{\mathrm{r}}$ denotes the running cost.
We impose the following assumptions, as in \cite{Desilles2019}.
\begin{assum}
	$J_{\mathrm{t}}$ is locally Lipschitz continuous on $\mathbb{R}^7$ with Lipschitz constant $L_{\mathrm{t}}(R)$ for every neighborhood $R \subset \mathbb{R}^7$.
	\label{assu:Lipschitz_Jfinal}
\end{assum}
\begin{assum}
	$J_{\mathrm{r}}$ is continuous on $\mathbb{R}^7 \times \mathcal{U}$. Moreover, $J_{\mathrm{r}}$ is locally Lipschitz continuous on the first variable with Lipschitz constant $L_{\mathrm{r}}(R)$ for every neighborhood $R \subset \mathbb{R}^7$.
	\label{assu:Lipschitz_Jrunning}
\end{assum}
\begin{rem}
    To ease notation, we omit the dependents on the neighborhood for the Lipschitz constants and instead assume the existence of a global Lipschitz constant $L_{\mathrm{t}}$ and $L_{\mathrm{r}}$, respectively.
\end{rem}
Next, we define an auxiliary state $z \in \mathbb{R}^p$ as:
\begin{equation}
\begin{cases}
	\dot{\mathbf{z}}(s) = -J_{\mathrm{r}}(\mathbf{r}(s), \mathbf{u}(s)), & \forall s \in [-t_f,0] \\
	\mathbf{z}(0) = z_0,
\end{cases}
\label{eqn: auxstate}
\end{equation}
where $z_0$ becomes an optimization parameter and $\mathbf{z} \in \mathbb{W}^{1,1}(\mathbb{R}^p)$. The auxiliary state captures the cumulative running cost and thus is treated as an additional state. In the same manner we previously ensured a trajectory, $\mathbf{r}$, stayed within the set $\mathcal{K}$, we bound $\mathbf{z}$ and ensure that the integrated running cost, added with the terminal cost, stays below some value $z_0$. To capture all possible trajectories, we introduce the set 
\begin{multline}
	\mathcal{Z}_{r_0, t_f, z_0}  \coloneqq \big\{(\mathbf{r}, \mathbf{u}, \mathbf{z}) : (\mathbf{r}, \mathbf{u}) \in \Pi_{r_0, t_f};\dot{\mathbf{z}}(s) = \\
	-J_{\mathrm{r}}(\mathbf{r}(s), \mathbf{u}(s)), 
	\forall s \in [-t_f,0];  \mathbf{z}(0) = z_0  \big\},
\end{multline}
and make the following assumption.
\begin{assum}
	For every $r \in \mathbb{R}^7$,the set
	\begin{equation*}
		\left\{\begin{bmatrix}
			f(r,u) \\
			-J_{\mathrm{r}}(r,u)
		\end{bmatrix} : u \in \mathcal{U}, \right\}
	\end{equation*}
	is a compact convex subset of $\mathbb{R}^{7} \times \mathbb{R}^p$.
	\label{assu:compactsubset}
\end{assum}
We now introduce the auxiliary value function $\vartheta$:
\begin{multline}
	\vartheta(r_0, t_f, z_0) \coloneqq \inf_{(\mathbf{r}, \textbf{u}, \mathbf{z}) \in \mathcal{Z}_{r_0, t_f, z_0}}  \Big\{ \\
	\bigvee_i \left[ J^i_{\mathrm{t}}(\mathbf{r}(0)) - \mathbf{z}^i(-t_f) \right]
	\bigvee
	\nu(\mathbf{r}(0))
	\bigvee
	\max_{s \in [-t_f, 0]} g(\mathbf{r}(s))
	\Big\},
	\label{eqn: aux value func}
\end{multline}
where $\bigvee_i x^i$ denotes the maximum element of the vector $x$. As with $\omega$, the term $\max_{s \in [-t_f, 0]} g(\mathbf{r}(s))$ and $\nu(\mathbf{r}(0))$ ensures that any trajectory $\mathbf{r}$ remains in $\mathcal{K}$ and terminates in $\mathcal{C}$. The additional term $J^i_{\mathrm{t}}(\mathbf{r}(0)) - \mathbf{z}^i(-t_f)$ ensures that the integrated running cost, $J_{\mathrm{r}}$, combined with the terminal cost, $J_{\mathrm{t}}$, never grows larger than $z_0$. Thus, in addition to ensuring that $(\mathbf{r},\mathbf{u})$ are admissible trajectory control pairs, the sub-zero level set of $\vartheta$ bounds the terminal and integrated running cost. Therefore,
\begin{equation}
	\vartheta(r_0, t_f, z_0)  \leq 0 \iff \Big[ \exists (\mathbf{r},\mathbf{u}) \in \Pi_{r_0, t_f}^{\mathcal{K}, \mathcal{C}}, J_{\mathrm{Bolza}}(\mathbf{r}, \mathbf{u}, t_f) \leq z_0 \Big].
\end{equation}
We are now in a position to introduce the generalized multi-objective optimal control problem for objective functions in Bolza form:
\begin{equation}
\begin{aligned}
& \minimize_{(r_0, t_f) \in \mathcal{I} \times [0, \infty)}
&  z_0 \\
& \text{subject to}
& \vartheta(r_0, t_f, z_0) \leq 0,
\end{aligned}
\label{eqn:genMoc}
\end{equation}
where $z_0$ represents an upper bound for the term $J_{\mathrm{Bolza}}(\mathbf{r}, \mathbf{u}, t_f)$, without explicit knowledge of $\mathbf{r}$ or $\mathbf{u}$.

The generalized value function can again be obtained as the unique continuous viscosity solution of a quasi-variational inequality
\begin{equation}
	\begin{cases}
		\begin{aligned}
			\max \big\{g(r)-&{}\vartheta(r, t, z), 
			\partial_{t} \vartheta + H(r, \nabla_{r} \vartheta, \nabla_{z} \vartheta)\big\} = 0\\
			& \text{for all} \: t \in [0, \infty), \:r \in \mathbb{R}^7, z \in \mathbb{R}^p,
		\end{aligned} \\		
		\begin{aligned}
			\vartheta(r, 0,z) = &{}
			\bigvee_i \left[ J^i_{\mathrm{t}}(r) - z^i \right]
			\bigvee
				\nu(r)
			\bigvee
				g(r) \\
			 &\text{for all} \: r \in \mathbb{R}^7,
		\end{aligned}
	\end{cases}
	\label{eqn:HJBgen}
\end{equation}
where the Hamiltonian is defined as
\begin{equation*}
	H(r, q_r, q_z) \coloneqq  \min_{u \in \mathcal{U}} \left(q_r^T f(r,u) - q_z^T J_{\mathrm{r}}(r, u)\right).
\end{equation*}
\begin{prop}
	The value function $\vartheta$ is Lipschitz continuous.
	\label{prop:vartheta-lips}
\end{prop}
The proof can be found in the Appendix. Proposition \ref{prop:vartheta-lips} can be used to show that a numerical solution of \eqref{eqn:HJBgen} (in the viscosity sense) can always be determined.

Under Assumptions \ref{assu:Lipschitz}, \ref{assu:Lipschitz_Jfinal}, \ref{assu:Lipschitz_Jrunning} and \ref{assu:compactsubset}, by Filippov's Theorem \cite{Liberzon2011}, the problem \eqref{eqn: aux value func} admits an optimal solution, which implies the existence of an admissible $\mathbf{r}$ and $\mathbf{z}$ \cite{Desilles2019}. This yields the following relationship due to \eqref{eqn: auxstate}
\begin{equation}
    \bigvee_i \left[ J^i_{\mathrm{t}}(\mathbf{r}(0)) - \mathbf{z}^i(-t_f) \right]= \\
    \bigvee_i \left[ J^i_{\mathrm{t}}(\mathbf{r}(0)) +  \int_{-t_f}^{0}  J^i_{\mathrm{r}}(\mathbf{r}(s), \mathbf{u}(s)) ds - z^i_0\right].
    \label{eqn:z-fullform}
\end{equation}

\section{Numerical Approximation and Results} \label{sec:numerics}
We will now discuss how the value functions can be obtained numerically, prior to discussing how the spacecraft trajectory design problem is solved. Following Proposition \ref{prop:omega-lips}, a numerical solution to \eqref{eqn:HJBgen} can be found. To this end, we employ the Level Set Methods toolbox of \cite{Mitchell2008}. For the computation of $\omega$, we use a Lax-Friedrich Hamiltonian 
\begin{equation}
	\mathcal{H}(r, p^-, p^+) = H(r, \frac{p^- + p^+}{2}) - \sum_{k=1}^7 \frac{\alpha_k}{2} (p^+ - p^-),
\end{equation}
where $p^+$ and $p^-$ are the right and left derivatives computed using an appropriate fifth-order weighted essentially non-oscillatory (WENO) scheme. The Lax-Friedrich Hamiltonian consists of an analytic expression of the Hamiltonian (derived previously), as well as a dissipation term, which is scaled by the dissipation coefficients $\alpha_k$. The dissipation coefficients $\alpha_k$ needs to satisfy
\begin{equation}
	\alpha_k \geq \left|\frac{\partial H}{\partial q_k} \right|.
\end{equation}
Since we need $\alpha_k$ to serve as an upper bound, we consider the control input that maximizes the Hamiltonian:
\begin{multline}
	 \left|\frac{\partial H}{\partial q_k} \right| =
	 \Big| - \frac{\partial C(r,q)}{\partial q_k} 
	 + \frac{\partial}{\partial q_k} \max \Big(
		q_7 \frac{T_{\max}}{v_{\mathrm{exhaust}}} \\
		+ \frac{T_{\max}}{m_0+\Delta m} \sqrt{q_4^2 + q_5^2 + q_6^2}, 0
		\Big) \Big|
\end{multline}
\begin{equation}
	\alpha_k = 
	\begin{cases}
		|v_\rho|				
		& k=1					\\ 
		|\frac{v_t}{\rho \sin{\psi}}	|		& k=2												\\ 
		|\frac{v_\perp}{\rho} 			|		& k=3											\\
		|a_\rho  - \max(0,\frac{T_{\max} q_3}{(m_0+\Delta m) \sqrt{q_3^2 + q_4^2 + q_5^2}}) |   & k=4    			\\
		|a_t  - \max(0, \frac{T_{\max} q_4}{(m_0+\Delta m) \sqrt{q_3^2 + q_4^2 + q_5^2}} )     |   & k=5  	\\
		|a_\perp - \max(0, \frac{T_{\max} q_5}{(m_0+\Delta m) \sqrt{q_3^2 + q_4^2 + q_5^2}})|	& k=6	\\
		\frac{T_{\max}}{v_{\mathrm{exhaust}}} & k=7,
	\end{cases}
\end{equation}
For a further discussion of the Lax-Friedrich Hamiltonian and WENO scheme, we refer to \cite{Osher2002}, while for a discussion of the convergence of $\omega$ and the derivation of a necessary Courant-Friedrichs-Lewy condition, we refer to \cite{Bokanowski2010, Hermosilla2018, Mitchell2008}.
\subsection{Implementation}
To illustrate the theoretical results of the previous sections, we consider a spacecraft on an initial near circular orbit around asteroid Castalia 4769. \textcolor{black}{The goal is to compute an efficient transfer trajectory that raises the spacecraft to a stable orbit at an altitude of $6117.5$ m above the asteroid. For the derivation of a stable orbit around Castalia 4769 we refer to \cite{Kulumani2016_cast} and references therein.} The gravity of Castalia 4769 was modeled by means of a spherical harmonic expansion as discussed in \cite{Hudson1994, Scheeres1987, Scheeres1996}. \textcolor{black}{Even though the proposed theoretical framework allows us to tackle problems of any state dimension, the available numerical tools and computational power limit us to only study the lower-dimensional planar case of the application. We, therefore, omit the states $\psi$ and $v_\perp$ to consider only 
\begin{equation}
    r   = \begin{bmatrix}
		\rho,
		\theta,
		v_\rho,
		v_t,
		\Delta m\end{bmatrix}^T \in \mathbb{R}^5.
\end{equation}}

To avoid ill-conditioning when solving the HJB equation, the state vector is normalized using the constants introduced in Table \ref{tab:normalization}. This results in the following dynamics:
\begin{equation}
	f(r,u)  = \begin{bmatrix}
		v_\rho															   \\ 
		\frac{v_t}{\rho} 												   \\ 
		a_\rho  + \frac{cT}{m_0+\Delta m} \cos{\alpha} \\
		a_t+ \frac{cT}{m_0+\Delta m} \sin{\alpha}       	\\
		- \frac{cT}{v_{\mathrm{exhaust}}}\end{bmatrix},
\end{equation}
where $c=T_\max \rho_0 /(m_0 V_0^2)$ is a normalization constant. $\rho_0$, $m_0$, and $v_0$ denote the initial radius, mass, and velocity of the initial orbit, respectively.
\begin{table}
\color{black}
\centering
\caption{Normalization}
\label{tab:normalization}
\begin{tabular}{|l|l|l|l|l|}
\hline
\textbf{Scale} & \multicolumn{4}{l|}{\textbf{Values}}               \\ \hline
Distance       & \multicolumn{4}{l|}{Initial radius $\rho_0$}           \\ \hline
Velocity       & \multicolumn{4}{l|}{$2$ m/s} 		 \\ \hline
Time           & \multicolumn{4}{l|}{$\rho_0/v_0$}                            \\ \hline
Mass           & \multicolumn{4}{l|}{$1$ kg}                             	   \\ \hline
Force          & \multicolumn{4}{l|}{Maximum thrust $T_\max$}     \\ \hline
\end{tabular}
\end{table}
The optimal control policy and trajectory $(\mathbf{r}, \mathbf{u}) \in \Pi_{r_0, t_f}^{\mathcal{K}, \mathcal{C}}$ can be constructed efficiently using the numerical approximation of $\omega$. For a given $N \in \mathbb{N}$ we consider the time step $h=\frac{1}{N}$ and a uniform grid of $[-t_f, 0]$ with spacing $s^k = \frac{k}{N}$. Let us define the state $\{r^k\}_{k=0} ^{N}$ and control $\{u^k\}_{k=0} ^{N-1}$ for the numerical approximation of the optimal trajectory and control policy. Setting $r_0$ as the initial orbit, we proceed by iteratively computing the control value
\begin{equation*}
	u^k(r^k) \in \argmin_{u \in \mathcal{U}}  \omega(r^k + h f(r^k,u), s^k) \bigvee g(r^k).
\end{equation*}
For a given $\omega(r^k, s^k)$, this is done by numerically taking the partial derivatives along each grid direction to estimate the costate vector, $q$, and then determining the optimal control values as the minimizer of the Hamiltonian, $H$. \textcolor{black}{After $u^k$ is determined we compute $r^{k+1}$ using the Matlab \texttt{ode113} function, a variable-order Adams-Bashforth-Moulton method of order 1 to 13 \cite{Shampine1997}, and increment $k$. For the implementation, we discretized the interval $[-t_f, 0]$ using $N=4000$ grid points.} 

\begin{figure}
	\centering
	\includegraphics[width=0.8\columnwidth]{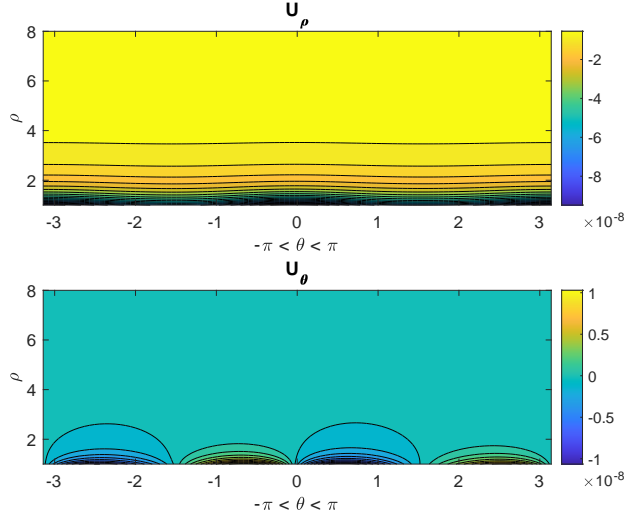}
	\caption{Gravitational acceleration comparing $U_{\rho}$ and $U_\theta$ around Castalia 4769.}
	\label{fig:grav}
\end{figure}
As shown in Figure \ref{fig:grav}, when considering orbits further than $4$ km away from the surface of the asteroid, the variation of the gravitational acceleration along $\theta$ becomes negligible. It is, therefore, possible to approximate the gravitational terms in spherical coordinates as 
\begin{align*}
	U_\rho(\rho,\theta) & \approx U_\rho(\rho) \\
	U_\theta(\rho,\theta) & \approx 0.
\end{align*}
Using this approximation makes $a_\rho$ and $a_t$ independent of $\theta$. This allows us to omit a grid dimension while numerically solving the quasi-variational inequality in Theorem \ref{thm:HJB}, greatly reducing the computational cost. \color{black} Thus the final set of states used for the computation of the value function is
\begin{equation}
    r   = \begin{bmatrix}
		\rho,
		v_\rho,
		v_t,
		\Delta m\end{bmatrix}^T \in \mathbb{R}^4.
\end{equation}
During the calculation of the optimal trajectory, $\theta$ can easily be reconstructed by forward integrating the dynamics at each time step $s^{k}$, i.e.,
\begin{equation}
    \theta^{k+1} = \theta^{k} + \int_{s^{k}}^{s^{k+1}} \frac{v_t^{k}}{\rho^{k}} ds.
\end{equation}
\color{black}

\subsection{Simulation results}
\begin{figure}
\color{black}
	\centering
	\includegraphics[width=0.8\columnwidth]{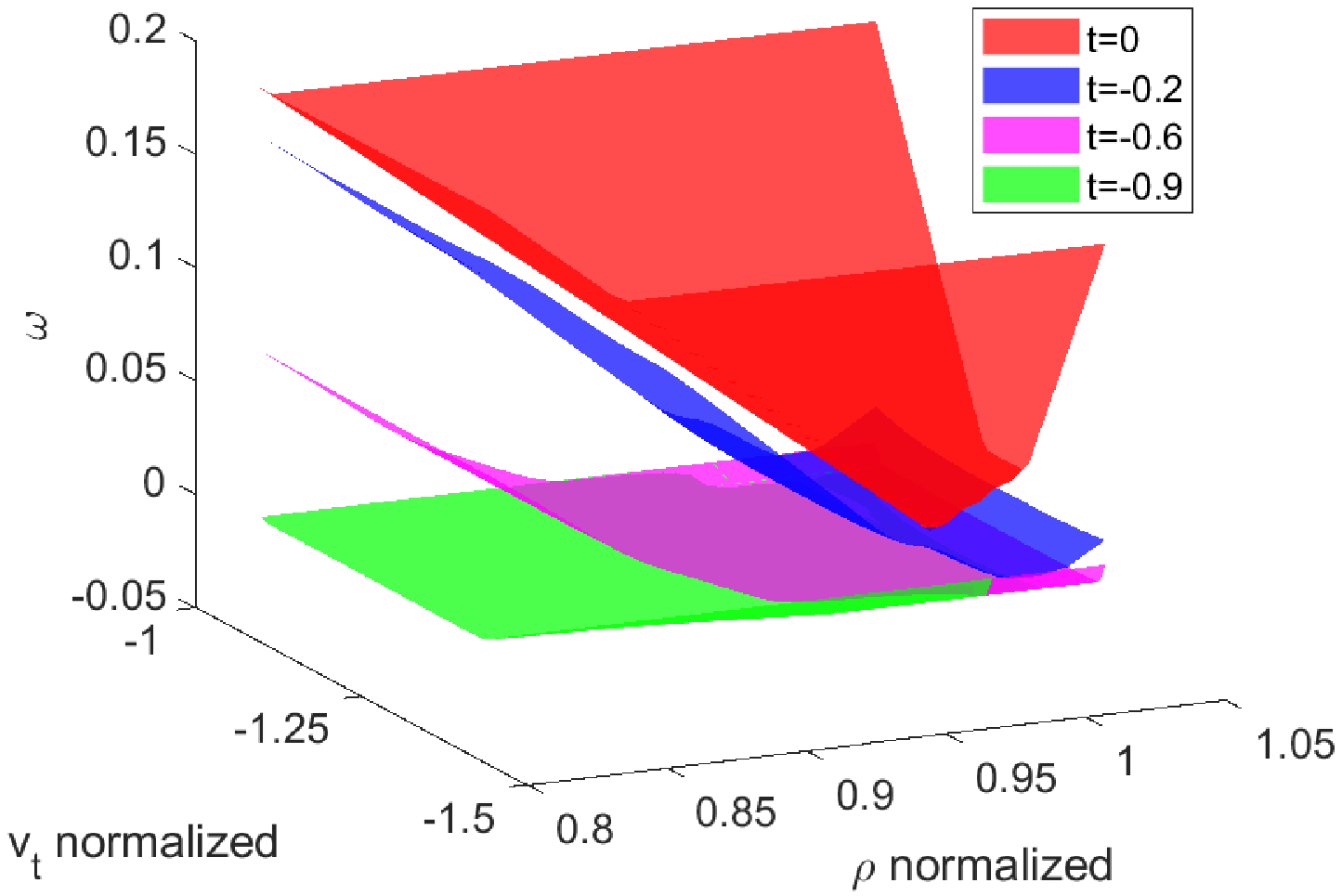}
	\caption{The evolution of the value function, $\omega$, projected in the $\rho$-$v_t$ plane for $v_\rho = 0$ and $\Delta m = 100$ g.}
	\label{fig:valuefunc}
\end{figure}

\begin{figure}
\color{black}
	\centering
	\includegraphics[width=0.8\columnwidth]{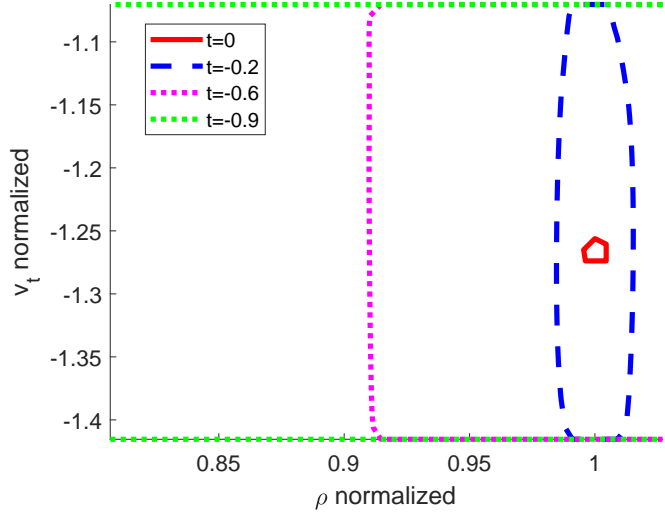}
	\caption{The evolution of the zero level set of the value function, $\omega$, projected in the $\rho$-$v_t$ plane for $v_\rho = 0$ and $\Delta m = 100$ g.}
	\label{fig:zerolevelset}
\end{figure}

\begin{figure}
\color{black}
	\centering
	\includegraphics[width=0.8\columnwidth]{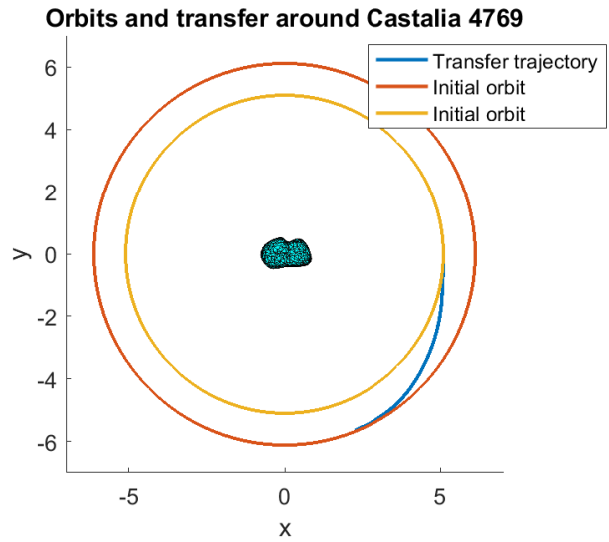}
	\caption{Initial orbit and transfer trajectory to a circular orbit $\approx1$ km further away from the asteroid.}
	\label{fig:traj}
\end{figure}

The spacecraft is modeled with $750$ kg of dry mass, $600$ mN of maximum thrust, and an exhaust velocity of $40$ km/s. \textcolor{black}{Using an initial orbit with radius $5.1$ km and tangential velocity of $-2.4$ m/s, we are able to compute the numerical approximation of $\omega$ using the spatial grid described in Table \ref{tab:gridding} in combination with a temporal grid using $N=1000$ grid points. The propagation of the zero level set over time is shown in Figure \ref{fig:valuefunc} and \ref{fig:zerolevelset}. Since the temporal grid for the calculation of the value function is more coarse than that used for the trajectory calculation, we need to interpolate the value function while computing the final trajectory.}
\begin{table}
\color{black}
\centering
\caption{Spatial grid configuration}
\label{tab:gridding}
\begin{tabular}{|l|c|c|c|c|}
\hline
\multicolumn{1}{|c|}{\textbf{}} & $\rho$ & $v_\rho$ & $v_t$ & $\Delta m$ \\ \hline
\textbf{Points}             & 50        & 40      & 40     & 32         \\ \hline
\textbf{Spacing}            & 0.0045    & 0.0416  & 0.0088  & 0.0067     \\ \hline
\textbf{Minimum}            & 0.8067     & -0.2495  & -1.4154 & -0.0533    \\ \hline
\textbf{Maximum}            & 1.0270     & 1.3722   & -1.0704 & 0.1533     \\ \hline
\end{tabular}
\end{table}
\textcolor{black}{The final computed trajectory, for an initial propellant mass of $24.89$ g and transfer time $2757$ s is shown in Figure \ref{fig:traj}.} The asteroid rendering for Figure \ref{fig:traj} was computed as in \cite{Kulumani2016_cast}.

\begin{figure}
\color{black}
	\centering
	\includegraphics[width=0.8\columnwidth]{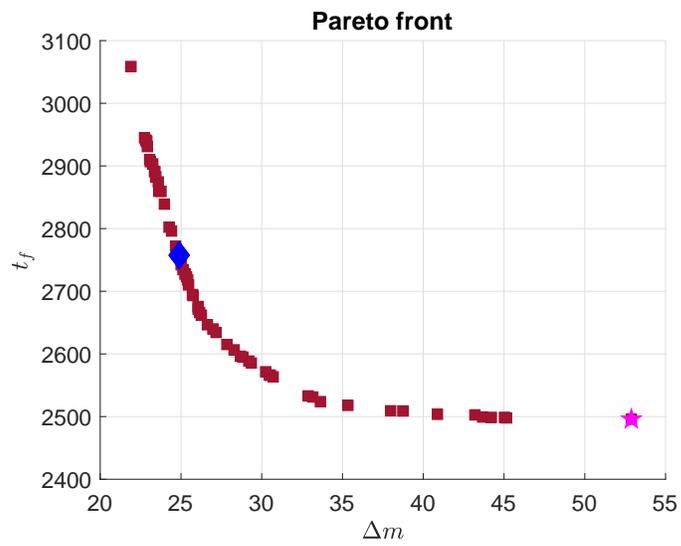}
	\caption{Pareto front of the objective functions $J_1=\Delta m$ in g and $J_2=t_f$ in seconds. The trajectory in Figure \ref{fig:traj} is derived using the point marked by the black diamond, while the time-optimal solution used in Figure  \ref{fig:control_figure} uses the point denoted by the magenta pentagram.}
	\label{fig:pareto}
\end{figure}

\textcolor{black}{The accuracy of the final orbit is within $41$ meters of the target orbit. Calculating $\omega$ took 9 hours using a 3 GHz 8-Core Intel Core i7-9700 processor running Matlab with an extension of the Level Set Methods toolbox \cite{Chen2016}. The run time and accuracy can be significantly improved upon when using optimized code such as \cite{Bokanowski2016, Fisac2016}. To show how the accuracy of the solution, the memory usage as well as the CPU time varies with the grid size, we recompute the value function and the trajectory with coarser spatial grids, as presented in Table \ref{tab:computation}. We use only single-precision arrays to store the value function, yet utilized double-precision arrays for all numerical calculations on the value function. The subscript \textit{final} denotes the values of the final point of the computed trajectory, i.e. $\mathbf{r(0)}$, while \textit{target} refers to the target orbit used to define $\mathcal{C}$ and subsequently initialize the computation of the value function.}
\begin{table}
\color{black}
\centering
\caption{Comparison of the accuracy, CPU time, and memory usage as the size of the grid changes}
\label{tab:computation}
\begin{tabular}{|l|c|c|c|}
\hline
\multicolumn{1}{|c|}{\textbf{}}                                 & Low                       & Medium            & High                  \\ \hline
Grid Size                                                       & [24 18 18 12]         & [32 24 24 16]         & [50 40 40 32]      \\ \hline
CPU Time                                                        & 20 min                & 44 min                & 9.28 hours              \\ \hline
Memory to store $\omega$                                        & 374 MB                & 1181 MB               & 10.25 GB               \\ \hline
$|\rho_{\mathrm{final}} - \rho_{\mathrm{target}}|$              & 82.72 m               & 64.72 m               & 40.83 m               \\ \hline
$|v_{\rho, \mathrm{final}} - v_{\rho, \mathrm{target}}|$        & 6.80e-5 $\frac{m}{s}$ & 4.26e-5 $\frac{m}{s}$ & 8.31e-7 $\frac{m}{s}$  \\ \hline
$|v_{t, \mathrm{final}} - v_{t, \mathrm{target}}|$              & 2.40e-4 $\frac{m}{s}$ & 4.01e-4 $\frac{m}{s}$ & 2.41e-4 $\frac{m}{s}$  \\ \hline
\end{tabular}
\end{table}
\textcolor{black}{Once $\omega$ is computed, it is incorporated into \eqref{eqn:newMoc}, which is solved using Matlab's \texttt{paretosearch} function. Solving the MOC problem took 120 seconds and the resulting Pareto front is shown in Figure \ref{fig:pareto}.}

\begin{figure}
        \color{black}
	\centering
	\includegraphics[width=0.8\columnwidth]{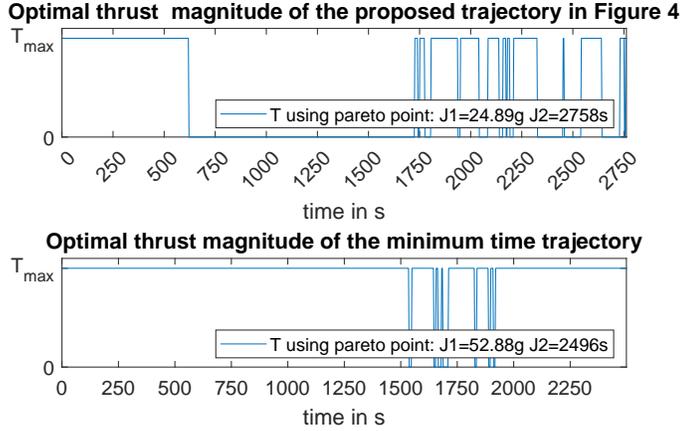}
	\caption{A comparison of the thrust magnitude of the smoothed optimal control policy of the time-optimal solution and the trajectory presented in Figure \ref{fig:traj}.}
	\label{fig:control_figure}
\end{figure}

\textcolor{black}{A comparison of the thrust magnitude of the smoothed optimal control policy is shown in Figure \ref{fig:control_figure}. As can be seen, the time-optimal solution uses near continuous thrust to reach the target, at the cost of using a large amount of fuel. The control policy of the trajectory presented in Figure \ref{fig:traj} meanwhile, has noticeable cruising phases where no fuel is consumed. As expected, the thrust magnitude of both policies follows a bang-bang structure.}

\begin{figure}
\color{black}
	\centering
	\includegraphics[width=0.8\columnwidth]{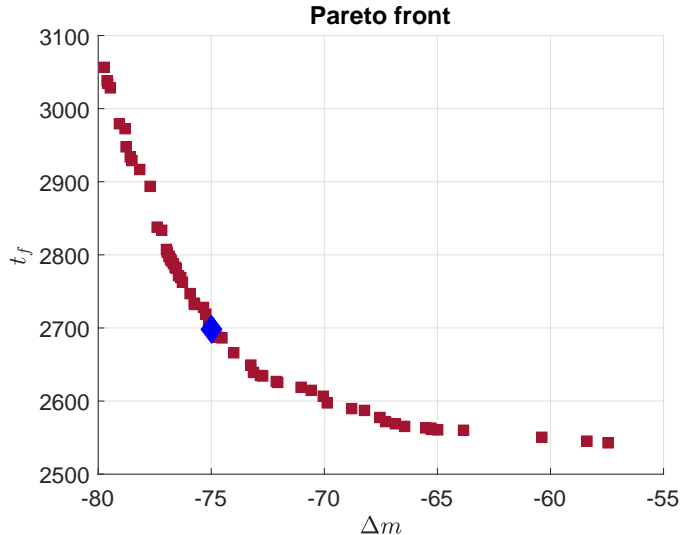}
	\caption{Pareto front of the objective functions $J_1=z$ in g and $J_2 = t_f$ in seconds, for the Bolza problem. The point used for trajectory calculations is marked by the black diamond.}
	\label{fig:pareto_vartheta}
\end{figure}

\textcolor{black}{To illustrate the results of problems in Bolza form, we consider the case of optimizing the remaining propellant in oppose to the initial propellant. Therefore, let us fix the initial propellant to $100$ g. We use a similar setup as in \ref{sec:numerics}, with the addition of the auxiliary state, $z$, defined as a terminal cost. We consider the uniform spaced grid over $r$ and $z$, defined in Table \ref{tab:gridding2}. Storing the value function using single precision requires 15.2 GB of memory.}
\begin{table}
\color{black}
\centering
\caption{Spatial grid configuration for the Bolza problem}
\label{tab:gridding2}
\begin{tabular}{|l|c|c|c|c|c|}
\hline
\multicolumn{1}{|c|}{\textbf{}} & $\rho$ & $v_\rho$ & $v_t$ & $\Delta m$ & $z$     \\ \hline
\textbf{Points}         & 32    & 24      & 24     & 32        & 32     \\ \hline
\textbf{Spacing}        & 0.007 & 0.069   & 0.016  & 0.0067    & 0.0067 \\ \hline
\textbf{Minimum}        & 0.805 & -0.277  & -1.446 & -0.0533   & -0.1533 \\ \hline
\textbf{Maximum}        & 1.029 & 1.317   & -1.068 & 0.1533    & 0.0533  \\ \hline
\end{tabular}
\end{table}
Since the objective is to maximize the remaining propellant, the optimization problem needs to minimize $-\Delta m$. Therefore, for a given final state $r_f \in \mathbb{R}^7$ and transfer time $t_f \in [0, +\infty)$, we define the cost functions as $J_1(r_f, t_f) \coloneqq -\Delta m$ and $J_2(r_f, t_f) \coloneqq t_f$, where $\Delta m$ denotes the $7$-th element of the state vector $r_f$ (the mass in our case). The 2-dimensional objective function $J : \mathbb{R}^7 \times [0,+\infty) \rightarrow \mathbb{R}^2$ can then be written as 
\begin{equation}
	J(r_f, t_f) \coloneqq \left[ J_1(r_f, t_f), J_2(r_f, t_f) \right]^T.
\end{equation}
The resulting Pareto front is shown in Figure \ref{fig:pareto_vartheta}. \textcolor{black}{Calculating the reachable set took approximately $32$ hours of CPU time. Using the reachable set, calculating the Pareto front took approximately $258$ seconds.} As expected, the Pareto front looks similar to that of Figure \ref{fig:pareto}, yet not identical due to numerical inaccuracy and change in the initial mass, resulting in modified dynamics.

\textcolor{black}{Using an initial propellant mass of $100$ g, z of $-75$ g, and transfer time of $2698$ s, the transfer orbit is computed in the same way as before. As expected, the transfer trajectory is similar to the trajectory shown in Figure \ref{fig:traj}, and the accuracy of the transfer orbit is within $66$ meters of the target orbit. For comparison to the first approach, the accuracy of the second approach is shown in Table \ref{tab:computation2}. From comparing the Pareto front, it can be seen that both methods have a minimum transfer time of around 2500 seconds as well as a minimum propellant requirement of just over 20 g.}

\begin{table}
\color{black}
\centering
\caption{Comparison of the accuracy, cpu time, and memory usage for the second formulation in Bolza form.}
\label{tab:computation2}
\begin{tabular}{|l|c|}
\hline
\multicolumn{1}{|c|}{\textbf{}}                                 & Bolza-formulation     \\ \hline
Grid Size                                                       & [32 24 24 32 32]      \\ \hline
CPU Time                                                        & 32 hours              \\ \hline
Memory to store $\omega$                                        & 15.2 GB               \\ \hline
$|\rho_{\mathrm{final}} - \rho_{\mathrm{target}}|$              & 66.14 m               \\ \hline
$|v_{\rho, \mathrm{final}} - v_{\rho, \mathrm{target}}|$        & 1.8e-6 $\frac{m}{s}$  \\ \hline
$|v_{t, \mathrm{final}} - v_{t, \mathrm{target}}|$              & 0.0018 $\frac{m}{s}$  \\ \hline
\end{tabular}
\end{table}

\section{Conclusion} \label{sec:conclusion}
We have presented a novel method of using the value function of a quasi-variational inequality to compute the decision space of multi-objective optimization problems. The feasibility and effectiveness of the proposed approach was demonstrated by applying it to the problem of low-thrust trajectory design. The approach is applicable to arbitrary multi-objective optimization problems where the control variable is required to lie within a reachable set.

Since the Hamiltonian becomes affine with respect to the thrust magnitude, once the thrust angles have been fixed, future research aims to exploit this fact by integrating classification based approaches \cite{Rubies-Royo2019a}. Furthermore, utilizing approximations of the reachable set as in \cite{Bansal2020a}, as well as decomposing the reachable sets as in \cite{Chen2018}, seems promising.
\bibliographystyle{plain}
\bibliography{bibliography}  
\appendix
\section{Proposition \ref{prop:bnd-r}}\label{appendix: bnd-r} 
\begin{prop}
	Under Assumption \ref{assu:Lipschitz}, any two trajectories $\mathbf{r}$ and $\mathbf{\hat{r}}$ reconstructed from $f$, with $\mathbf{r}(-t_f) = r_0$ and $\mathbf{\hat{r}}(-t_f) = \hat{r}_0$, respectively, are such that $\big| \big| \mathbf{r}(\tau) - \mathbf{\hat{r}}(\tau) \big| \big| \leq \big| \big| r_0 - \hat{r}_0 \big| \big| e^{(t_f+\tau) L_f}$ for all $\tau \in [-t_f, 0]$.
	\label{prop:bnd-r}
\end{prop}
\begin{pf}
Let $r_0, \hat{r}_0 \in \mathbb{R}^7$ be two initial states, and $t_f \in [0,\infty)$. For the same $t_f$, we choose two trajectory control pairs
$(\mathbf{r}, \mathbf{u}) \in \Pi_{r_0, t_f}$ and $(\mathbf{\hat{r}}, \mathbf{\hat{u}}) \in \Pi_{\hat{r}_0, t_f}$. Then by Carath{\'e}odory's existence of solutions \cite{Sastry1999}, the following relation holds:
\begin{align*}
	\big| \big| \mathbf{r}(-t) - \mathbf{\hat{r}}(-t) \big| \big| & \leq 
	\big| \big| r_0 - \hat{r}_0 \big| \big| + \int_{-t_f}^{-t} \big| \big| f(\mathbf{r}(s),\mathbf{u}(s)) - f(\mathbf{\hat{r}}(s), \mathbf{\hat{u}}(s)) \big| \big| ds \\
	& \leq \big| \big| r_0 - \hat{r}_0 \big| \big| + L_f \int_{-t_f}^{-t} \big| \big| \mathbf{r}(s) - \mathbf{\hat{r}}(s) \big| \big| ds \\
	& \leq \big| \big| r_0 - \hat{r}_0 \big| \big| e^{(t_f-t) L_f},
\end{align*}
where the second inequality is due to Assumption \ref{assu:Lipschitz}, while the last inequality is due to the Bellman-Gronwall Lemma \cite{Sastry1999}.
\end{pf}
\section{Proof of Proposition \ref{prop:vartheta-lips}} \label{appendix:vartheta-lips}
\begin{pf}
Fix $(r_0, z_0), (\hat{r}_0, \hat{z}_0)\in \mathbb{R}^7 \times \mathbb{R}^p$, $t_f \in [0,\infty)$ and let $\epsilon > 0$. We choose $(\hat{\mathbf{r}}, \hat{\mathbf{u}}, \hat{\mathbf{z}}) \in \mathcal{Z}_{\hat{r}_0,t_f, \hat{z}_0,}$ such that
\begin{equation*}
	\vartheta(\hat{r}_0, t_f, \hat{z}_0) 
	\geq \bigvee_i \left[ J^i_{\mathrm{t}}(\mathbf{\hat{r}}(0)) - \mathbf{\hat{z}}^i(-t_f) \right] \bigvee \nu(\mathbf{\hat{r}}(0)) \bigvee \max_{s \in [-t_f, 0]} g(\mathbf{\hat{r}}(s)) - \epsilon.
\end{equation*}
By definition of $\vartheta$, for any $(\mathbf{r}, \mathbf{u}) \in \Pi_{{r}_0,{t}}$, this yields the following relation
\begin{multline*}
	\vartheta(r_0, t_f, z_0) - \vartheta(\hat{r}_0, t_f, \hat{z}_0) 
	\leq \bigvee_i \left[ J^i_{\mathrm{t}}(\mathbf{r}(0)) - \mathbf{z}^i(-t_f) \right] \bigvee \nu(\mathbf{r}(0))
		\bigvee
		\max_{s \in [-t_f, 0]} g(\mathbf{r}(s)) \\
		- \bigvee_i \left[ J^i_{\mathrm{t}}(\mathbf{\hat{r}}(0)) - \mathbf{\hat{z}}^i(-t_f) \right] \bigvee \nu(\mathbf{\hat{r}}(0)) \bigvee \max_{s \in [-t_f, 0]} g(\mathbf{\hat{r}}(s)) + \epsilon.
\end{multline*}
Let $\kappa \in [-t_f, 0]$ be such that \begin{equation*}
    g(\mathbf{r}(\kappa)) = \max_{s \in [-t_f, 0]} g(\mathbf{r}(s)).
\end{equation*}
We then have
\begin{multline*}
	\vartheta(r_0, t_f, z_0) - \vartheta(\hat{r}_0, t_f, \hat{z}_0) 
	\leq \bigvee_i \left[ J^i_{\mathrm{t}}(\mathbf{r}(0)) - \mathbf{z}^i(-t_f) \right] \bigvee \nu(\mathbf{r}(0)) \bigvee g(\mathbf{r}(\kappa)) \\
		- \bigvee_i \left[ J^i_{\mathrm{t}}(\mathbf{\hat{r}}(0)) - \mathbf{\hat{z}}^i(-t_f) \right] \bigvee \nu(\mathbf{\hat{r}}(0)) \bigvee g(\mathbf{\hat{r}}(\kappa))+ \epsilon.
\end{multline*}
Using Proposition \ref{prop:bnd-r}, we distinguish the following cases.

\textit{Case A}: $g(\mathbf{r}(\kappa)) \geq \bigvee_i \left[ J^i_{\mathrm{t}}(\mathbf{r}(0)) - \mathbf{z}^i(-t_f) \right] \bigvee \nu(\mathbf{r}(0))$
\begin{multline*}
	\vartheta(r_0, t_f, z_0) - \vartheta(\hat{r}_0, t_f, \hat{z}_0) 
	\leq \\
	g(\mathbf{r}(\kappa)) - \bigvee_i \left[ J^i_{\mathrm{t}}(\mathbf{\hat{r}}(0)) - \mathbf{\hat{z}}^i(-t_f) \right] \bigvee \nu(\mathbf{\hat{r}}(0)) \bigvee g(\mathbf{\hat{r}}(\kappa))+ \epsilon \\
	\leq g(\mathbf{r}(\kappa)) - g(\mathbf{\hat{r}}(\kappa)) + \epsilon \leq L_g e^{(t_f+\kappa) L_f} \big| \big| r_0 - \hat{r}_0 \big| \big| + \epsilon,
\end{multline*}
where the last inequality is due to the fact that $g$ is Lipschitz continuous.

\textit{Case B}: $\nu(\mathbf{r}(0)) \geq \bigvee_i \left[ J^i_{\mathrm{t}}(\mathbf{r}(0)) - \mathbf{z}^i(-t_f) \right] \bigvee g(\mathbf{r}(\kappa))$
\begin{multline*}
	\vartheta(r_0, t_f, z_0) - \vartheta(\hat{r}_0, t_f, \hat{z}_0)
	\leq \\
	\nu(\mathbf{r}(0)) - \bigvee_i \left[ J^i_{\mathrm{t}}(\mathbf{\hat{r}}(0)) - \mathbf{\hat{z}}^i(-t_f) \right] \bigvee \nu(\mathbf{\hat{r}}(0)) \bigvee g(\mathbf{\hat{r}}(\kappa))+ \epsilon \\
	\leq \nu(\mathbf{r}(0)) - \nu(\mathbf{\hat{r}}(0)) + \epsilon \leq L_{\nu} e^{t_f L_f} \big| \big| r_0 - \hat{r}_0 \big| \big| + \epsilon
\end{multline*}

\textit{Case C}: $ \bigvee_i \left[ J^i_{\mathrm{t}}(\mathbf{{r}}(0)) - \mathbf{{z}}^i(-t_f) \right] \geq g(\mathbf{r}(\kappa)) \bigvee \nu(\mathbf{r}(0))$

Recall \eqref{eqn:z-fullform}, then under Assumption \ref{assu:Lipschitz}, any two trajectories $\mathbf{z}$ and $\mathbf{\hat{z}}$ reconstructed from $J_{\mathrm{r}}$ with $z_0$ and $\hat{z}_0$, respectively, are bounded within a given time interval $[-t, 0]$. To see this,
\begin{align*}
	\big| \big| \mathbf{z}(-t) - \mathbf{\hat{z}}(-t) \big| \big| 
	& \leq \big| \big| z_0 - \hat{z}_0 \big| \big| + \int_{-t}^0 \big| \big| J_{\mathrm{r}}(\mathbf{r}(s), \mathbf{u}(s)) - J_{\mathrm{r}}(\mathbf{\hat{r}}(s), \mathbf{\hat{u}}(s)) \big| \big| ds \\
	& \leq \big| \big| z_0 - \hat{z}_0 \big| \big| + L_{\mathrm{r}} \int_{-t}^0 \big| \big| \mathbf{r}(s) - \mathbf{\hat{r}}(s)\big| \big| ds\\
	& \leq \big| \big| z_0 - \hat{z}_0 \big| \big| + L_{\mathrm{r}} \int_{-t}^0 \big| \big| r_0 - \hat{r}_0 \big| \big| e^{(t_f+s) L_f} ds \\
	& \leq \big| \big| z_0 - \hat{z}_0 \big| \big| + \big| \big| r_0 - \hat{r}_0 \big| \big|  L_{\mathrm{r}} e^{t_f L_f} \frac{1-e^{-t L_f}}{L_f},
\end{align*}
where the third inequity is due to Proposition \ref{prop:bnd-r}, and the last one follows by performing the integration.

Next, let $j \in [1, \ldots ,p]$ be such that
\begin{equation*}
    J^j_{\mathrm{t}}(\mathbf{{r}}(0)) - \mathbf{{z}}^j(-t_f) = \bigvee_i \left[ J^i_{\mathrm{t}}(\mathbf{{r}}(0)) - \mathbf{{z}}^i(-t_f)\right].
\end{equation*}
Then it follows, that
\begin{align*}
	\vartheta(r_0, t_f, z_0) - \vartheta(\hat{r}_0, t_f, \hat{z}_0)
	& \leq \bigvee_i \left[ J^i_{\mathrm{t}}(\mathbf{{r}}(0)) - \mathbf{{z}}^i(-t_f) \right]\\
	& - \bigvee_i \left[ J^i_{\mathrm{t}}(\mathbf{\hat{r}}(0)) - \mathbf{\hat{z}}^i(-t_f)\right]\bigvee \nu(\mathbf{\hat{r}}(0)) \bigvee g(\mathbf{\hat{r}}(\kappa))+ \epsilon \\
	& \leq \bigvee_i \left[ J^i_{\mathrm{t}}(\mathbf{{r}}(0)) - \mathbf{{z}}^i(-t_f) \right] \\ & -\bigvee_i \left[ J^i_{\mathrm{t}}(\mathbf{\hat{r}}(0)) - \mathbf{\hat{z}}^i(-t_f) \right]+ \epsilon \\
	& \leq \left[ J^j_{\mathrm{t}}(\mathbf{{r}}(0)) - \mathbf{z}^j(-t_f) \right]  - \left[ J^j_{\mathrm{t}}(\mathbf{\hat{r}}(0)) - \mathbf{\hat{z}}^j(-t_f) \right]+ \epsilon \\
	& \leq \left[ J^j_{\mathrm{t}}(\mathbf{{r}}(0)) - J^j_{\mathrm{t}}(\mathbf{\hat{r}}(0)) \right] - \left[ \mathbf{z}^j(-t_f) - \mathbf{\hat{z}}^j(-t_f)\right] + \epsilon
\end{align*}
By Proposition \ref{prop:bnd-r} and under Assumption \ref{assu:Lipschitz_Jfinal}
\begin{equation*}
    \left[ J^j_{\mathrm{t}}(\mathbf{{r}}(0)) - J^j_{\mathrm{t}}(\mathbf{\hat{r}}(0)) \right] \leq L_{\mathrm{t}} \big| \big| r_0 - \hat{r}_0 \big| \big| e^{t_f L_f}.
\end{equation*}
Finally, this yields the relationship
\begin{multline*}
    \left[ J^j_{\mathrm{t}}(\mathbf{{r}}(0)) - J^j_{\mathrm{t}}(\mathbf{\hat{r}}(0)) \right] - \left[ \mathbf{z}^j(-t_f) - \mathbf{\hat{z}}^j(-t_f)\right]  + \epsilon \\
    \leq \big| \big| z_0 - \hat{z}_0 \big| \big| + \big| \big| r_0 - \hat{r}_0 \big| \big|  \Big[L_{\mathrm{t}} e^{t_f L_f} + L_{\mathrm{r}} \frac{e^{t_f L_f}-1}{L_f}\Big] + \epsilon.
\end{multline*}
Thus in every case, on an interval $[0, t_f]$ there exists a set of constants $C_r$ and $C_z$, such that
\begin{equation*}
	\vartheta(r_0, t_f, z_0) - \vartheta(\hat{r}_0, t_f, \hat{z}_0) \\
	\leq C_r \big| \big| r_0 - \hat{r}_0 \big| \big| + C_z \big| \big| z_0 - \hat{z}_0 \big| \big| + \epsilon
\end{equation*}
The same argument conducted with $(r_0, t_f, z)$ and $(\hat{r}_0, t_f, \hat{z})$ reversed establishes that
\begin{equation*}
	\vartheta(\hat{r}_0, t_f, \hat{z}_0) - \vartheta(r_0, t_f, z_0) \\
	\leq C_r \big| \big| r_0 - \hat{r}_0 \big| \big| + C_z \big| \big| z_0 - \hat{z}_0 \big| \big| + \epsilon.
\end{equation*}
Since $\epsilon$ is arbitrary, we conclude that 
\begin{equation*}
	\big| \big| \vartheta(\hat{r}_0, t_f, \hat{z}_0) - \vartheta(r_0, t_f, z_0) \big| \big| \\
	\leq C_r \big| \big| r_0 - \hat{r}_0 \big| \big| + C_z \big| \big| z_0 - \hat{z}_0 \big| \big|,
\end{equation*}
thus concluding the proof.
\end{pf}
The proof for $\omega$ is similar to that of $\vartheta$ and will therefore be omitted in the interest of space.

\color{black}
\section{Derivation of the optimal thrust angles}
Notice that since the applied thrust, $T$, is always positive, the term
\begin{equation*}
    \left(q_4 \cos{\alpha} +\sin{\alpha} \left(q_5 \sin{\delta} +q_6 \cos{\delta}\right) \right)
\end{equation*}
in the Hamiltonian can be minimized separately from $T$. 

To this end, we introduce the auxiliary variables 
\begin{align*}
	\chi(\delta) \coloneqq &{} \sqrt{q_5^2 + q_6^2} \cos{(\delta - \arctan{\frac{q_5}{q_6}})}, \\
	A(\delta) \coloneqq & \sqrt{q_4^2 + \chi(\delta)^2},
\end{align*}
using the trigonometric identity
\begin{equation*}
a \cos{x} + b \sin{x} = R \cos{(x - \arctan{\frac{b}{a}})},
\end{equation*}
with $R = \sqrt{a^2 + b^2}$. 
First optimizing over $\alpha$, and subsequently over $\delta$ (notice that this sequential minimization is exact since $A(\delta) \geq 0$) results in
\begin{multline*}
\min_{\alpha, \delta \in [-\pi, \pi]\times[-\frac{\pi}{2}, \frac{\pi}{2}]}\left(
	q_4 \cos{\alpha} +
	\sin{\alpha} \left(
		q_5 \sin{\delta} +
		q_6 \cos{\delta}
		\right) \right)  \\
	= \min_{\delta \in [-\frac{\pi}{2}, \frac{\pi}{2}]} A(\delta) \min_{\alpha \in [-\pi, \pi]} \cos{(\alpha - \arctan{\frac{\chi(\delta)}{q_4}})}.
\end{multline*}
Thus, the optimal thrust angles are given by
\begin{equation*}
	\alpha^*(\delta) \coloneqq \pm \pi + \arctan{\frac{\chi(\delta)}{q_4}}.
\end{equation*}
Since $\cos{(\alpha^*(\delta) - \arctan{\frac{\chi(\delta)}{q_4}})} = -1$, after applying $\alpha^*(\delta)$, it follows that 
\begin{equation*}
	\delta^* \in \argmin_{\delta \in [-\frac{\pi}{2}, \frac{\pi}{2}]} - A(\delta) = \arctan{\frac{q_5}{q_6}} \pm \pi.
\end{equation*}
Subsequently,
\begin{equation}
     q_4 \cos{\alpha^*} +
	\sin{\alpha^*} \left(
		q_5 \sin{\delta^*} +
		q_6 \cos{\delta^*}
		\right) \\
	= -\sqrt{q_4^2 + q_5^2 + q_6^2}.
	\label{eqn: angles}
\end{equation}
Substituting \eqref{eqn: angles} into \eqref{eqn:Hamiltonian_bfr_optangle} results in \eqref{eqn:Hamiltonian_aftr_optangle} which depends in an affine fashion on $T$.
\end{document}